\documentclass[traditabstract]{aa}
\usepackage{natbib,epsfig,graphicx,ulem}
\usepackage{amsmath,amsfonts,amssymb}
\usepackage{txfonts}
\usepackage{longtable}
\usepackage{soul,color}
\usepackage[breaklinks,colorlinks,citecolor=blue]{hyperref}

\newcommand{\Euclid}{{\it Euclid}}

\newcommand{\sigmae}{\sigma_\epsilon}

\newcommand{\OmegaM}{\Omega_{\rm m}}

\begin{document}

\title{Constraining cosmology with shear peak statistics: tomographic analysis}

\titlerunning{Constraining cosmology with shear peak
  statistics} \authorrunning{Martinet et al.}

\author{Nicolas Martinet\inst{1,3} \and James G. Bartlett\inst{2,3} \and Alina Kiessling\inst{3} \and Barbara Sartoris\inst{4,5}}
\offprints{Nicolas Martinet, \email{martinet@iap.fr}}

\institute{UPMC Universit\'e Paris 06, UMR~7095, Institut
  d'Astrophysique de Paris, 98bis Bd Arago, F-75014, Paris, France
  \and 
  APC, AstroParticule et Cosmologie, Universit\'e Paris Diderot, CNRS/IN2P3, CEA/lrfu, Observatoire de Paris, Sorbonne Paris Cit\'e, 10, rue Alice Domon et L\'eonie Duquet, 75205 Paris Cedex 13, France
  \and
  Jet Propulsion Laboratory, California Institute of Technology, 4800
Oak Grove Drive, Pasadena, California, U.S.A.
  \and
  Dipartimento di Fisica, Sezione di Astronomia, Universit\`a di Trieste, Via Tiepolo 11, I-34143 Trieste, Italy
  \and
  INAF/Osservatorio Astronomico di Trieste, Via Tiepolo 11, I-34143 Trieste, Italy}

\setcounter{page}{1}

\abstract {The abundance of peaks in weak gravitational lensing maps is a potentially powerful cosmological tool, complementary to measurements of the shear power spectrum. We study peaks detected directly in shear maps, rather than convergence maps, an approach that has the advantage of working directly with the observable quantity, the galaxy ellipticity catalog. Using large numbers of numerical simulations to accurately predict the abundance of peaks and their covariance, we quantify the cosmological constraints attainable by a large-area survey similar to that expected from the \Euclid\ mission, focusing on the density parameter, $\OmegaM$, and on the power spectrum normalization, $\sigma_8$, for illustration.  We present a tomographic peak counting method that improves the conditional (marginal) constraints by a factor of 1.2 (2) over those from a two-dimensional (i.e., non-tomographic) peak-count analysis. We find that peak statistics provide constraints an order of magnitude less accurate than those from the cluster sample in the ideal situation of a perfectly known observable-mass relation; however, when the scaling relation is not known a priori, the shear-peak constraints are twice as strong and orthogonal to the cluster constraints, highlighting the value of using both clusters and shear-peak statistics.
}
\keywords{weak lensing, shear, cosmology}

\maketitle

\section{Introduction}
\label{sec:intro}
Weak gravitational lensing (WL) is a powerful probe of large-scale structure, dark matter, and dark energy \citep[e.g.,][]{Bartelmann2001}.  In particular, cosmic shear surveys have demonstrated their ability to constrain cosmological parameters \citep[e.g.,][]{Massey2007, Heymans2012}, and the potential of large-area shear surveys covering thousands of square degrees to improve cosmological constraints to percent level accuracies is the prime motivation for ambitious programs like the on-going Dark Energy Survey (DES)\footnote{{\tt http://www.darkenergysurvey.org}}, and Stage IV experiments \citep{albrecht2006, albrecht2009} like the Large Synoptic Survey Telescope\footnote{{\tt http://www.lsst.org}} \citep[LSST,][]{LSSTSciBook2009}, the Wide Field Infrared Survey Telescope (WFIRST)\footnote{{\tt http://wfirst.gsfc.nasa.gov}} \citep{WFIRST-AFTA2013}, and the \Euclid\footnote{{\tt http://www.cosmos.esa.int/web/euclid}}\ \citep{euclidrb} missions.  

The shear correlation function (equivalently, power spectrum), a measure of the second moment of the mass distribution, is the standard tool for analyzing WL surveys.  Other statistical measures of shear maps incorporating higher order moments are possible, and they become increasingly attractive in light of the significant gain in signal-to-noise expected from the planned large WL surveys.  

In this paper, we consider the statistics of peaks in a shear map (or shear catalog) as a cosmological probe.  We define shear peaks by filtering the map with an aperture designed to detect localized projected mass concentrations, such as galaxy clusters.  Indeed, WL surveys can be used to detect galaxy clusters, and the cluster counts then used as a cosmological probe \citep{Kruse1999, Marian2006}.  Projection effects, however, severely limit the purity of cluster samples defined through WL, despite attempts at constructing optimal filters, because many peaks result from the alignment of small systems along the line of sight \citep{White2002,Hamana2004,Hennawi2005}.

An alternative is to abandon the correspondence between shear peaks and clusters and simply use the statistics of peaks to characterize the projected mass distribution \citep{Reblinsky1999}.  This is the approach we adopt in the present work.  These general shear peaks do not necessarily have any meaning as physical objects, being a combination of real clusters and chance alignments. Their abundance, however, like clusters, is sensitive to the underlying cosmology \citep{Jain2000,Wang2009,Dietrich+10,Kratochvil2010}.

One disadvantage of this approach is that we do not posses a simple analytic form for the abundance of peaks as a function of cosmological parameters.  This is in contrast to the situation with clusters, where practical expressions do exist for the mass function that greatly facilitate the theoretical prediction of cluster abundance and exploration of parameter space \citep{PS1974,Jenkins2001,Tinker2008}.  We must instead resort to N-body simulations to predict WL peak abundance, and we require large suites of simulations to explore the parameter space.  

Such studies have been undertaken by several authors in recent years \citep{Dietrich+10,Kratochvil2010,Yang2011,Hilbert+12,Marian2012,Marian2013}.  In this paper, we perform a Fisher analysis of the constraints from peak counts in the context of upcoming Stage IV dark energy surveys, comparing our results to constraints expected from cluster counts.  We employ large suites of independent N-body simulations to mitigate what has been an important  limitation of previous studies, and we work directly with shear measurements, rather than reconstructed convergence maps.

The {\small SUNGLASS} code \citep{Kiessling2011} is a rapid simulation tool based on line-of-sight integration through N-body boxes to calculate the WL field.  Its speed allows us to generate large numbers of simulated shear maps, and hence determine peak abundance as a function of cosmological parameters and its variance.  In particular, we are able to accurately  calculate the derivative of peak abundance with respect to the parameters needed for the Fisher matrix. It is important to note that while many previous works have employed N-body simulations for similar analyses, they have relied on statistically shifted maps generated from a limited number of simulations \citep[e.g.,][]{Dietrich+10,Kratochvil2010,Yang2011,Hilbert+12,Marian2012,Marian2013}. By contrast, the maps in this work are truly independent, with each map generated from a separate N-body realization.  

We work directly with ellipticity measurements, an unbiased estimator of the shear in the WL regime \citep[e.g.,][]{Dietrich+10,Maturi+11,Hamana2012,Hilbert+12}, rather than convergence maps that have been used in several previous studies \citep[e.g.,][]{Kratochvil2010, Yang2011, Marian2012, Marian2013}.  Convergence is not the direct observable, but must be reconstructed from shear measurements.  Our approach avoids the complexity added by this inversion. In addition, the use of shear greatly simplifies the nature of map noise, which is non-trivial to estimate in the case of the reconstructed convergence.

Finally, we apply tomography to the peak statistics by dividing the lensed background galaxies into redshift bins.  This offers a two-fold advantage; first it allows us to remove foreground galaxies ($z\leq0.5$) that tend to dilute the shear signal, although the mass distribution below this redshift still contributes to the statistics measured using only the higher redshift galaxies.  Second, we can detect shear peaks in different redshift planes and examine the statistics both within and between planes.  As with the shear correlation function, the additional radial information significantly increases precision on cosmological constraints. Building on the work of \citet{Hennawi2005} and \citet{Dietrich+10}, we quantify the constraining power of tomographic shear peak statistics for Stage IV dark energy missions, such as $\Euclid$.

We begin by describing our WL simulations in Section~\ref{sec:simulations}.  The peak detection procedure and its application to the simulations are detailed in Section~\ref{sec:peak}.   In Section~\ref{sec:stats} we examine peak statistics and their use as a cosmological probe.  Section~\ref{sec:tomography} extends the approach to tomography.  We conclude with a final discussion and comment on future directions in Section~\ref{sec:discussion}.  Throughout the paper, for concreteness, we consider the specific case of a survey similar to that of the \Euclid\  mission with a fiducial flat $\Lambda$CDM cosmology specified by \{$\Omega_{\rm M}$, $\Omega_{\Lambda}$, $\Omega_{\rm b}$, $h$, $\sigma_8$, $n_{\rm S}$\}$=$\{0.272, 0.728, 0.0449, 0.71, 0.809, 1.000\} \citep[ e.g.,][]{Hinshaw2013}.
 

\section{Weak lensing simulations}
\label{sec:simulations}
We employ the Fisher formalism to asses the cosmological constraints expected from peak counts in Stage IV dark energy experiments, taking as typical characteristics those of the \Euclid\  \citep{euclidrb} mission.  \Euclid\  will survey 15,000 deg.$^2$ in three infrared bands (Y, J, and H) and a single, wide optical filter (combined riz bands), the latter with a point spread function (PSF) of 0.1\,arcsec.  The survey reaches a mean source galaxy density usable for lensing of 30 gals.~arcmin$^{-2}$ with a median redshift of $z=0.8$, following the redshift distribution
\begin{equation}
\label{eq:distriz}
 P(z) = \frac{3}{2{z_{0}}^3}z^2e^{-(\frac{z}{z_0})^{1.5}},
\end{equation}
with $z_0=0.7$.

Calculation of the Fisher matrix requires the derivative of the mean peak counts with respect to cosmological parameters, evaluated at the fiducial model.  We also need the covariance of the peak counts about their mean in this model.  Since we do not posses an analytical expression for the WL peak abundance, we must use simulations to calculate the expected peak counts.  We need enough simulations to accurately determine the mean peak counts for each parameter variation and to determine their covariance in the fiducial model.  This is non-trivial as much of the signal comes from non-linear scales in the mass distribution, which can only be properly modeled by N-body simulations.  

Simulation speed is therefore essential.  In our study we use the {\small SUNGLASS} pipeline developed by \citet{Kiessling2011}. We give a brief description of the {\small SUNGLASS} pipeline here but for a full prescription of how the {\small SUNGLASS} WL shear and convergence catalogs are generated, see \citet{Kiessling2011}. For a given set of parameters, {\small SUNGLASS} first generates an N-body realization with the {\small GADGET-2} code~\cite{springel2005}. These simulations are performed with 512$^3$ paticles in a 512h$^{-1}$.Mpc box and the light cone is 100 square degrees and goes out to a redshift of $z=2.0$. The convergence and shear of each mass point are calculated along the line of sight at multiple lensing source planes. This relies on the applicability of the Born approximation in the WL regime, i.e., that the mean path of the light bundle from a distant object remains adequately straight in the presence of lensing.  The integration is much quicker than ray tracing through the simulation box, and allows the production of large suites of simulated WL observations. The shear and convergence is then interpolated on to the individual particles in the light cone, providing a highly sampled catalog of shear and convergence along the line of sight.

The speed of {\small SUNGLASS} enables us to produce many independent realizations, an improvement over previous studies that had to resort to shuffling the results from single realizations. The final WL catalogs are constructed by down-sampling the highly sampled {\small SUNGLASS} shear and convergence catalogs to 30 galaxies per square arc minute using the source galaxy redshift distribution of Eq.~(\ref{eq:distriz}), and assuming that the galaxies trace the dark matter exactly.
To model the shape noise arising from intrinsic galaxy ellipticity, we add a random ellipticity to each source galaxy in the simulated shear catalogs according to a Gaussian distribution with zero mean and standard deviation $\sigmae$.  In practice, we use a Box-Muller Gaussian random number generator, drawing two uniform numbers $(x,y)$ over the
interval $[0,1)$ to obtain two Gaussian random numbers $(\epsilon_1,\epsilon_2)$ that we add, respectively, to the
shear components, $\gamma_1$ and $\gamma_2$:
\begin{equation}
 \epsilon_1=\frac{\sigma_\epsilon}{\sqrt{2}}\sqrt{-2\ln{(x)}}\cos{(2\pi y)}
\end{equation}
\begin{equation}
 \epsilon_2=\frac{\sigma_\epsilon}{\sqrt{2}}\sqrt{-2\ln{(x)}}\sin{(2\pi y)}.
\end{equation}
Null values in the first step are replaced by arbitrarily chosen values. 

This procedure corresponds to drawing a tangential ellipticity from a Gaussian of zero mean and dispersion $\sigma_\epsilon$, which we take to be $\sigma_\epsilon=0.3$ \citep[e.g.,][]{leauthaud2007}.  We note that the shear components, $\gamma_1$ and $\gamma_2$, are typically only a small percent of the noise, $\epsilon_1$ and $\epsilon_2$.  Under the assumption that the noise is uncorrelated (i.e., in the absence of important intrinsic alignment effects), the mean ellipticity averaged over a number of source galaxies approaches the shear signal.  We will discuss this fundamental hypothesis in Section~\ref{sec:tomography}.



\section{Peak detection}
\label{sec:peak}
\subsection{Method}
To identify peaks in a shear catalog, we employ aperture mass filtering \citep{Schneider98,Bartelmann2001} with an outer annulus to remove the integration constant, thereby resolving the finite space inversion problem.  This technique does not return the true mass within the aperture, for example, when centered on a cluster; for that, one would have to know the true shape of the mass distribution as seen through the filter.  By adopting a mass distribution, such as the NFW profile \citep{NFWpro}, one can determine the true mass and also find clusters in WL surveys \citep[e.g.,][]{Marian2012, Marian2013}.  Our goal, however, is not to measure the true mass of physical objects, but to compare the number of peaks expected in a WL survey for different cosmologies.


The aperture mass can be calculated either from the convergence field, $\kappa$, or the shear field, $\gamma$.  In the former case, the aperture mass is calculated by integrating the convergence within the aperture centered  at 
position $\vec{\theta}_0$ (a two-dimensional vector in the plane of the sky),
\begin{equation}
  M_{\rm ap}(\vec{\theta}_0)=\int d^2 \theta \; U(|\vec{\theta}-\vec{\theta}_0|)\kappa(\vec{\theta}),
\end{equation}
where $U$ is a filter chosen to best fit the lens mass density profile. 

In a WL survey, the observable is actually the source galaxy ellipticity, $\epsilon$.  It is related to the reduced shear, $g$, by
\begin{equation}
\label{eq:redshear}
\langle\epsilon\rangle=g=\frac{\gamma}{1-\kappa} \approx \gamma,
\end{equation}
which tends to the shear, $\gamma$, in the weak lensing regime where $(\kappa,\gamma)<<(1,1)$.  The indicated average is over random orientations of intrinsic galaxy ellipticity. 
A WL survey thus directly measures shear, $\gamma$, rather than the convergence.  Working directly with the observable quantity, $\gamma$, we avoid the non-trivial step of integrating the shear to obtain the convergence field, which is a derived quantity.  

Shear and convergence are two mathematically distinct, although related, quantities, the first being a spinor of spin two and the second a scalar on the sphere. To adapt the expression for the aperture mass to the case of shear, we first define the scalar tangential shear, $\gamma_t$, for a galaxy image that accounts for both components of the shear ($\gamma_1$ and $\gamma_2$),
\begin{equation}
\gamma_{\rm t}(\vec{\theta},\vec{\theta}_0)=-\left[\gamma_1(\vec{\theta}) \cos 2\phi(\vec{\theta},\vec{\theta}_0) + \gamma_2(\vec{\theta}) \sin 2\phi(\vec{\theta},\vec{\theta}_0) \right],
\end{equation}
where $\phi$ is the angle giving the position of the galaxy image ($\vec{\theta}$) relative to an arbitrary fixed axis running through the center of the aperture, at position $\vec{\theta}_0$, in the image plane; this fixed axis defines a local cartesian coordinate system in the plane of the sky with origin positioned on the filter center.  The aperture mass equation can then be rewritten in terms of the tangential shear and the new filter function $Q$:
\begin{eqnarray}
Q(|\vec{\theta}|) & = &  \frac{2}{\theta^2} \int_0^{\theta} d^2 \theta' \theta' \; U (|\vec{\theta}'|)  - U(|\vec{\theta}|), \\
M_{\rm ap}(\vec{\theta}_0) & = & \int d^2 \theta \; Q(|\vec{\theta}-\vec{\theta_0}|)\gamma_{\rm t} (\vec{\theta},\vec{\theta}_0).
\end{eqnarray}

The convergence has been used in several previous weak lensing peak studies \cite[e.g.,][]{Kratochvil2010,Yang2011,Marian2012,Marian2013}, although recent studies increasingly work directly with the shear \citep[e.g.,][]{Dietrich+10,Maturi+11,Hamana2012,Hilbert+12}, which is also the approach adopted in this paper. 

We use the following weight functions, $U$ for convergence and $Q$ for shear, appropriate for a circular aperture \cite[see][]{Bartelmann2001}:  
\begin{equation}
U(\theta)= \left\{\begin{array}{rl}
\frac{1}{\pi\theta_1^{2}} &\mbox{if $0<\theta<\theta_1$} \\
\frac{1}{\pi(\theta_2^{2}-\theta_1^{2})} &\mbox{if $\theta_1<\theta<\theta_2$} \\
0 &\mbox{elsewhere}
\end{array}\right.
\end{equation}
\begin{equation}
Q(\theta)= \left\{\begin{array}{rl}
\frac{\theta_2^{2}}{\pi(\theta_2^{2}-\theta_1^{2})\theta^{2}} &\mbox{if $\theta_1<\theta<\theta_2$} \\
0 &\mbox{elsewhere}
\end{array}\right. .
\end{equation}
In practice, the integral over the aperture weight function becomes a sum weighted by the number density of galaxy images, $n$,

\begin{equation}
  M_{\rm ap}(\vec{\theta}_0) =\frac{1}{n} \sum_i U(|\vec{\theta}_i-\vec{\theta}_0|)\kappa (\vec{\theta}_i)
\end{equation}

\begin{equation}
  M_{\rm ap}(\vec{\theta}_0)=\frac{1}{n}\sum_i Q(|\vec{\theta}_i-\vec{\theta}_0|)\gamma_{{\rm t}}(\vec{\theta}_i, \vec{\theta}_0),
\end{equation}
where $\gamma_{t}(\vec{\theta}_i,\vec{\theta}_0)$ and
$\kappa(\vec{\theta}_i)$ are the tangential shear and the convergence
of the image at $\vec{\theta}_i$ relative to the point $\theta_0$. 

By expressing the equations in terms of discrete sums, we explicitly account for the sampling inherent in the observations, i.e., we only measure the shear where there is a source galaxy.  Moreover,  we use the actual number density of galaxies in the aperture, rather than a fixed, average value. The link from simulations to observations is trivial, as it is sufficient to replace the tangential shear, $\gamma_{t}$, in the last equation by the tangential component of the ellipticity, $\epsilon_{t}$. This is the principal interest of using shear instead of convergence. 

The aperture mass is convenient because in the case of the shear, it permits simple calculation of its variance due to the intrinsic ellipticity of the source galaxies, $\sigma_{\epsilon}$:
\begin{equation}
  \sigma(M_{\rm ap})=\frac{\sigma_{\epsilon}}{\sqrt{2}n}\left(\sum_{i}{Q^{2}(|\vec{\theta}_i-\vec{\theta}_0|)}\right)^{1/2}.
\end{equation}
This allows us to define a local noise level and peak detection threshold, which is another strong argument in favor of using shear peaks rather than convergence.  We then define peak amplitude as 
\begin{equation}
\label{eq:shearSN}
 \Gamma(\vec{\theta}_0)=\frac{M_{\rm ap}}{\sigma(M_{\rm ap})}=\frac{\sum_{i}{Q(|\vec{\theta}_i-\vec{\theta}_0|)\gamma_{t}(\vec{\theta}_i,\vec{\theta}_0)}}{\frac{\sigma_{\epsilon}}{\sqrt{2}}(\sum_{i}{Q^{2}(|\vec{\theta}_i-\vec{\theta}_0|)})^{1/2}}.
\end{equation}

The amplitude of a peak is independent of the normalization of the weight function, but does depend on the number of galaxies in the aperture through both the sums in numerator and denominator.  Peak amplitude thus varies as the square root of the number of galaxies in the aperture.  Intrinsic galaxy ellipticity also affects the signal-to-noise value, although it does not change the relative intensity of the peaks.  As mentioned, we take $\sigma_\epsilon=0.3$ \citep{leauthaud2007}.
 
A critical point is the size and shape of the aperture.  The chosen shape will preferentially select a specific form of structure, such as  clusters or filaments, while the size will favor one cluster size over others.  We must also adapt the aperture to include enough galaxies to optimize the signal-to-noise over the random shape noise (this point will be discussed in the section on tomography).  

For simplicity, we adopt a radially symmetric aperture of fixed angular radius. The inner radius is set to $\theta_1 = 3.5\arcmin$, corresponding to the typical size of a cluster at redshift $z=0.3$, where the contribution to shear peaks is the most important \citep{Dietrich+10}, to exclude any contamination by cluster galaxies and the strong lensing regime, while the outer radius is set to $\theta_2 = 10\arcmin$, which is roughly the limit of the lensing effect at this redshift \citep[see][]{Hamana2012}.  In practice, we have found that the mass inside a given aperture strongly depends on the inner radius.  In a future work, we plan to use a set of  aperture sizes to extract information from different scales. One could also use an adaptive matched filter to preferentially select galaxy clusters \citep[e.g.,][]{Marian2012}.  This is not our goal in this first study, and we leave the identification of an optimal filter to a future work.  Finally, peaks are selected to be larger than all their neighbors in a radius equal to that of the aperture. Peaks situated at less than $\theta_2$ from the map edges are discarded as they are not computed in the proper aperture.

\section{Peak statistics}
\label{sec:stats}
We first present our statistical methodology and results from a non-tomographic analysis of the peak counts.  Section~\ref{sec:tomography} then extends the analysis to tomographic peak counts.

\subsection{Method}
We implemented two statistical measures: a $\chi^{2}$ test and the Fisher information matrix, both defined over bins of peak height.
We chose our bins to include the same number of peaks based on the mean peak counts in our fiducial cosmology.  This bin size is then maintained for the other cosmologies.  Bin widths are given in Table~\ref{tab:bins}, along with the number of peaks for one fiducial realization.

\begin{table}[ht!]
  \caption{Bin widths and peak counts for one fiducial realization. The second column gives the lower bound on the peak signal-to-noise for each bin and the third the total number of peaks in the bin; for example, the first line reads: in bin 1 there are 70 peaks with signal-to-noise between 3.0 and 3.2.}
\centering
\begin{tabular}{ccc}
  \hline
  \hline bin number & bin size  &  N \\
 \hline 

1  &3.0  & 70  \\
2  &3.2  & 72  \\
3  &3.4  & 65  \\
4  &3.5  & 67  \\
5  &3.7  & 72  \\
6  &3.9  & 74  \\
7  &4.2  & 69  \\
8  &4.5  & 77  \\
9  &5.0  & 66  \\
10  &5.7  & 90  \\

  \hline 
  \hline

\end{tabular}
\label{tab:bins}
\end{table} 

\begin{figure}[ht!]
 \centering
 \includegraphics[width=0.35\textwidth,clip,angle=270]{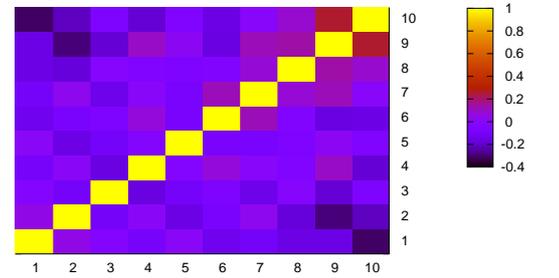} 
 \caption{Correlation matrix of the shear peak distribution for a 100 square-degree field, calculated by averaging over 150 realizations of the fiducial cosmology (see Eq.~\ref{eq:cov}).  This is the covariance matrix normalized to unit diagonal. The $x$ and $y$-axis values are the peak height bin numbers, defined in Table~\ref{tab:bins}.}
  \label{fig:correl}
\end{figure}

Let $N_{i,r}$ be the number of peaks in bin $i$ of realization $r$ for a given cosmology, and $R$ be the total number of realizations of this cosmology.  Defining $\langle N\rangle_i$ as the mean number of peaks in bin $i$, averaged over all $R$ realizations, we calculate the covariance matrix of the binned peak counts as 
\begin{equation}\label{eq:cov}
 C_{i,j}=\left\{\begin{array}{rl}
\frac{1}{R-2}\sum_{r}^{R}{(N_{i,r}- \langle N \rangle _i)(N_{j,r}- \langle N \rangle _j)} &\mbox{if $i \neq j$} \\
\frac{1}{R-1}\sum_{r}^{R}{(N_{i,r}- \langle N \rangle _i)(N_{j,r}- \langle N \rangle _j)} &\mbox{if $i = j$} 
\end{array}\right. ,
\end{equation} 
using $R=150$ independent realizations of the fiducial cosmology. Each of these realizations corresponds to a lightcone of 100deg$^2$, and we subsequently normalize the covariance matrix to an area of 15,000deg$^2$ (e.g., the useable extragalactic sky and \Euclid\ target). Figure~\ref{fig:correl} shows the correlation matrix, i.e., the covariance matrix normalized to unity along the diagonal.  We see that the correlation between bins is less than 20\% except for the higher signal-to-noise bins where it can reach up to 40\%. It seems reasonable that the stronger peaks would be more correlated between bins, with signal being dominated and seen by successive source planes, while the lower signal-to-noise peaks would be more affected by noise variations and projections between source planes. This agrees with the fact that high signal-to-noise peaks mostly correspond to galaxy clusters, while low signal-to-noise peaks are dominated by projections of large-scale structure and noise, as shown by \citet{Maturi+11}.

This covariance matrix is then used to compute either the $\chi^2$ (Eq.~\ref{eq:chi}) or the Fisher matrix (Eq.~\ref{eq:fish}).  The $\chi^2$ is expressed as
\begin{equation}
\label{eq:chi}
 \chi^2(r,f') = \sum_{i,j}{(N_{i,r}^{f'}- \langle N^f \rangle _{i})(C_{i,j}^f)^{-1}(N_{j,r}^{f'}- \langle N^f \rangle _{j})},
\end{equation} 
where $f$ represents the fiducial cosmology and $f'$ a reference cosmology.  We note that $f'$ can be the same as $f$ if we wish to compare one realization of a cosmology to all the other realizations of the same cosmology. This allows us, in particular, to test whether the $\chi^2$ variable is actually distributed according to a $\chi^2$-distribution.  The Fisher matrix is given by
\begin{equation}
\label{eq:fish}
F_{p_{\rm a}, {\rm p_b}} = \sum_{i,j}{\frac{\partial{\langle N \rangle _{i}}}{\partial{p_{\rm a}}}\widehat{({C}_{i,j}^{f})^{-1}}\frac{\partial{\langle N \rangle _{j}}}{\partial{p_{\rm b}}}},
\end{equation} 
where $p_{\rm a}$ and $p_{\rm b}$ are two cosmological parameters.  We note that this expression is easily interpretable only in the case of Gaussian distributed bin counts, i.e., as the number of peaks becomes large.  Cosmological parameter constraints are then obtained by inverting the Fisher matrix,
\begin{equation}
 C_{p_{\rm a},p_{\rm b}} = (F)^{-1}_{p_{\rm a},p_{\rm b}}.
\end{equation} 

Following \citet{Hartlap+07}, we use the unbiased estimator, $\widehat{C^{-1}}$, for the inverse covariance matrix. Under the assumption of Gaussian errors and independent data vectors, it is related to the inverse of the estimated covariance matrix (Eq.~\ref{eq:cov}) through the number of realizations, $R$, and the number of degrees-of-freedom, $D$, as
\begin{equation}
\widehat{C^{-1}} = \frac{R-D-2}{R-1}C^{-1},
\end{equation}
where in our case $D$ is the number of peak-height bins. This correction will thus be more important in our tomographic analysis where we build the covariance matrix through the assembled peak distributions of several redshift slices.

The derivative of the peak counts with respect to cosmological parameters averaged over all the realizations
$r$ is given by
\begin{equation}
\label{eq:deriv}
\frac{\partial{\langle N \rangle _{i}}}{\partial{p_{\rm a}}} = \frac{1}{R}\sum_{r=1}^{R}{\frac{N_{i.r}(p_{\rm a}+\Delta p_{\rm a})-N_{i.r}(p_{\rm a}-\Delta p_{\rm a})}{2\Delta p_{\rm a}}},
\end{equation}
where $\Delta p_{\rm a}$ is the variation of the cosmological parameter $p_{\rm a}$.  
This calculation requires a sufficient number, $R$, of numerical simulations of each cosmology to accurately determine the derivatives.  For a single parameter variation, we used 250 cosmological simulations (150 for the fiducial cosmology and 50 for a variation of $+\Delta p_{\rm a}$ and 50 for a variation of $-\Delta p_{\rm a}$).

\begin{figure}
\centering
\includegraphics[width=0.3\textwidth,clip,angle=270]{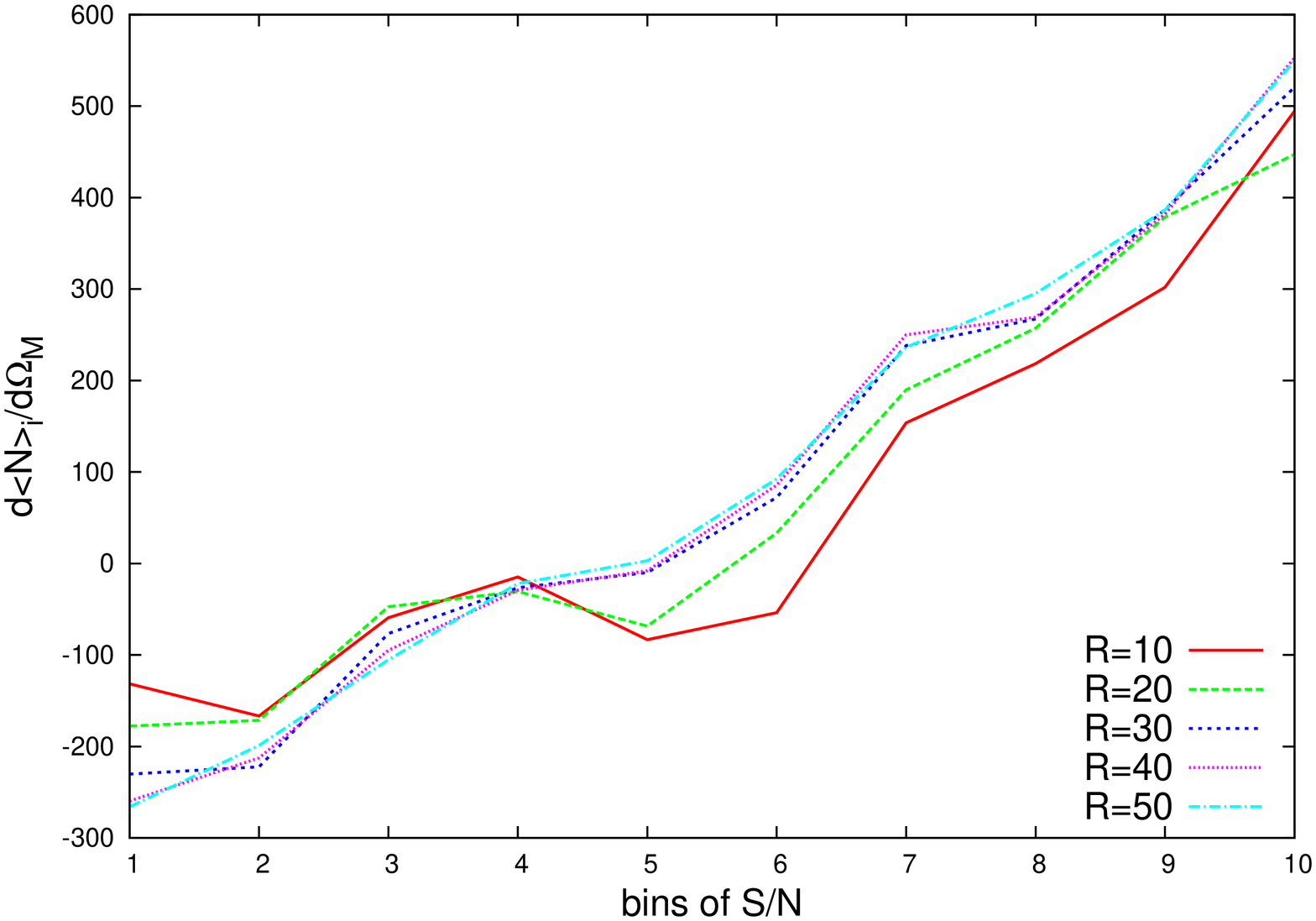}
\includegraphics[width=0.3\textwidth,clip,angle=270]{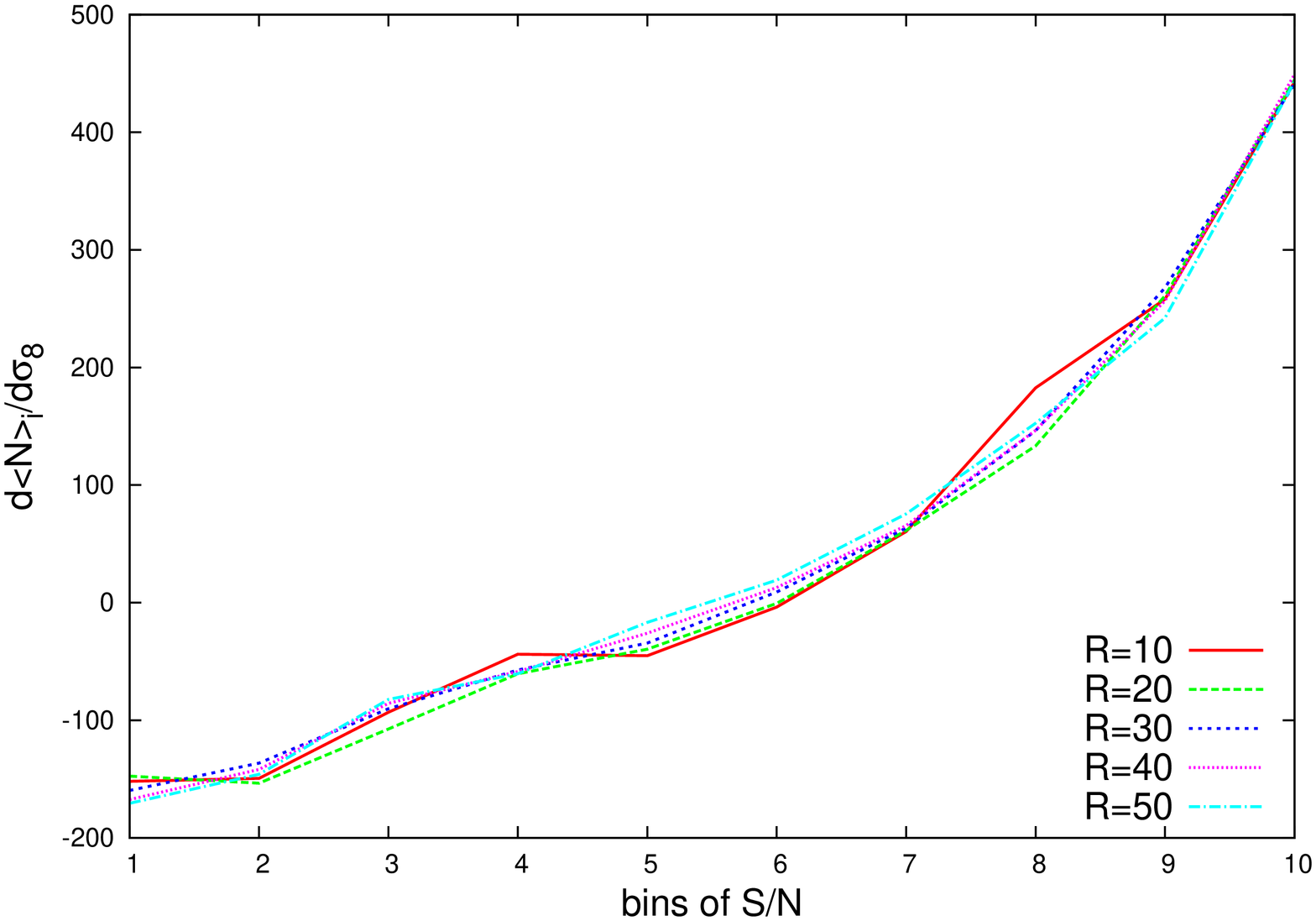}
\caption{Derivatives of the peak counts with respect to $\OmegaM$ (top) and $\sigma_8$ (bottom) as defined in Eq.~(\ref{eq:deriv}) as a function of the number of realizations of the modified cosmologies.  Results for $R=(10,20,30,40,50)$ correspond, respectively, to the red, green, blue, purple and cyan curves. There is little change beyond $R=30$.} 
\label{fig:deriv}
\end{figure}

To determine if this is sufficient, we perform a convergence test on numbers of realizations by comparing the derivatives of the peak counts for increasing numbers of realizations of the modified cosmologies. In Fig.~\ref{fig:deriv} we show the derivatives with respect to $\OmegaM$ and $\sigma_8$ when varying the number of realizations from 10 to 50 in increments of 10 realizations. We note that the derivatives do not significantly evolve beyond 30 realizations, justifying our choice of $R=50$ for the modified cosmologies.

We use a larger number of realizations of the fiducial cosmology because the covariance matrix is computed for that model, while only the mean peak counts are required for the other cosmologies. \citet{Taylor+14} calculated the accuracy of the covariance matrix given the number of realizations and of degrees of freedom of the data vectors. Following their Eq.~(13), we estimate the precision of our covariance matrix to be better than $\sim 13\%$ with our 150 realizations. The additive factor of $2\nu^{-2}$, where $\nu$ is the desired accuracy, in the required number of realizations limits in practice the achievable accuracy on the covariance matrix; sub-percent accuracy, for instance, would demand at least 40,000 realizations.  Our choice of 150 seems reasonable for the present test-study, but this issue calls for further attention and presents a crucial difficulty for many dark energy probes based on large-scale structure, such as cosmic shear.

\subsection{Results}
As an illustration of peak count statistics, we study achievable constraints on two cosmological parameters: the total matter density, $\OmegaM$, and the present-day linear matter power spectrum normalization, $\sigma_8$. In the standard $\Lambda$CDM model, these are well constrained by cosmic microwave background (CMB) observations \citep{Hinshaw2013, planck2013-xvi}.  Methods measuring their values at low redshifts, such as peak counts or other gravitational lensing observations, are useful to search for extensions of this simple model.  The power spectrum normalization, $\sigma_8$, is a good example: differences between values obtained from the CMB and low redshift methods could indicate the need for a non-minimal neutrino mass \citep[e.g.,][]{planck2013-xx, rozo2013, battye2014}.

Each parameter was varied by 10\% from its fiducial value given in Table~\ref{tab:param} to define a reference cosmology.  We generated 150 realizations of the fiducial cosmology and 50 realizations of each reference cosmology.  When  varying one parameter, the other remains at its fiducial value. However, when varying the matter density parameter, $\Omega_M$, the dark energy density parameter, $\Omega_{\Lambda}$, was also adjusted in order to maintain a flat Universe, $\Omega_M + \Omega_{\Lambda} = 1$.

\begin{table}[ht!]
  \caption{Cosmological parameter values for the fiducial and reference cosmologies. The quantity $R$ is the number of realizations of each cosmology. When varying one parameter, the other remains fixed at its fiducial value.}
\centering
\begin{tabular}{cccc}
  \hline
  \hline             &  $\Omega_{\rm M}$ & $\sigma_8$  & $R$\\ 
 \hline 
 	  fiducial   & 0.272     &  0.809  & 150 \\
 	  low        & 0.245     &  0.728  & 50  \\
 	  high       & 0.299     &  0.890  & 50  \\
  \hline 
  \hline

\end{tabular}
\label{tab:param}
\end{table}

\subsubsection{chi-squared distribution}
\begin{figure}[h]
\centering
\includegraphics[width=0.3\textwidth,clip,angle=270]{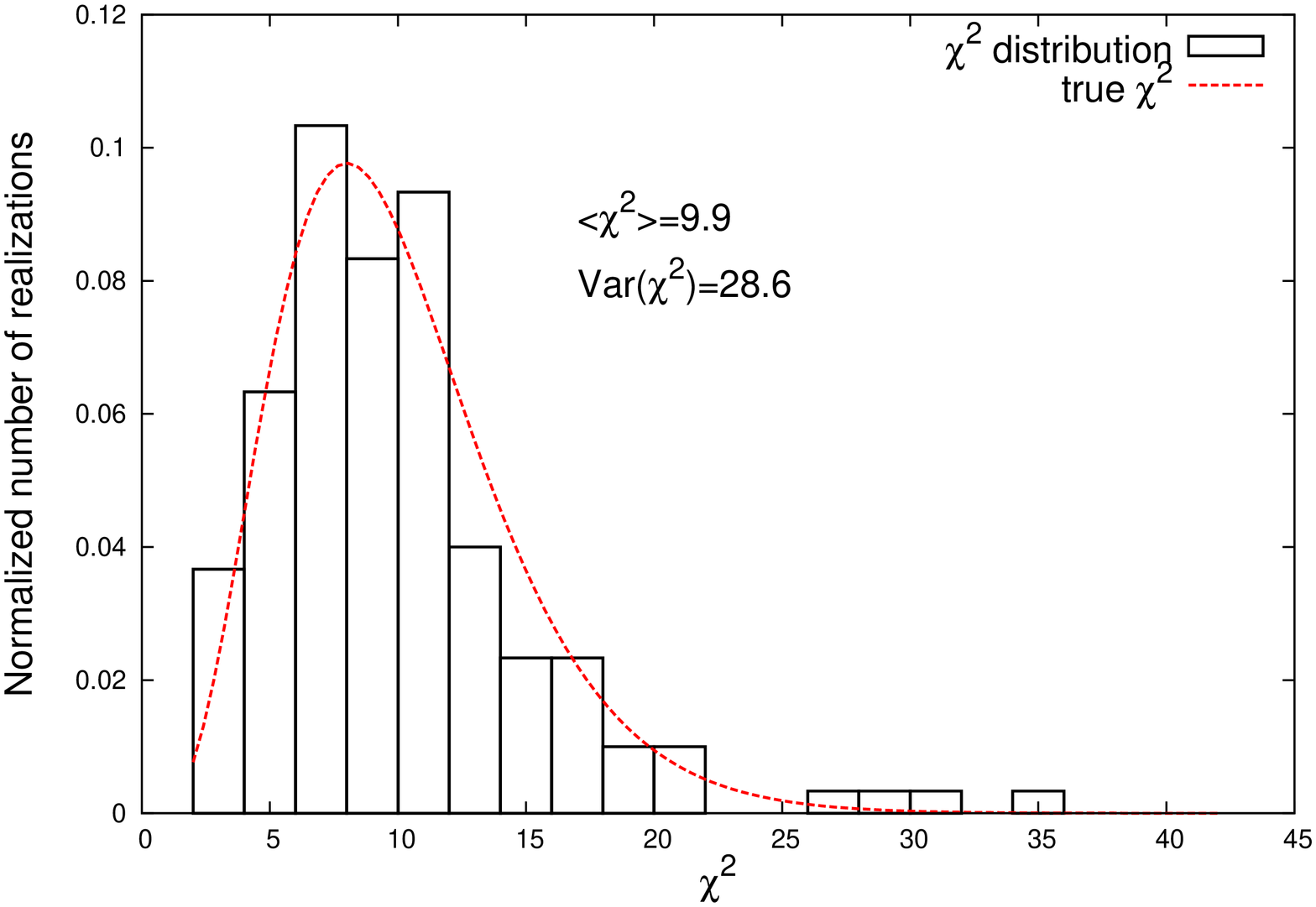} \\
\includegraphics[width=0.3\textwidth,clip,angle=270]{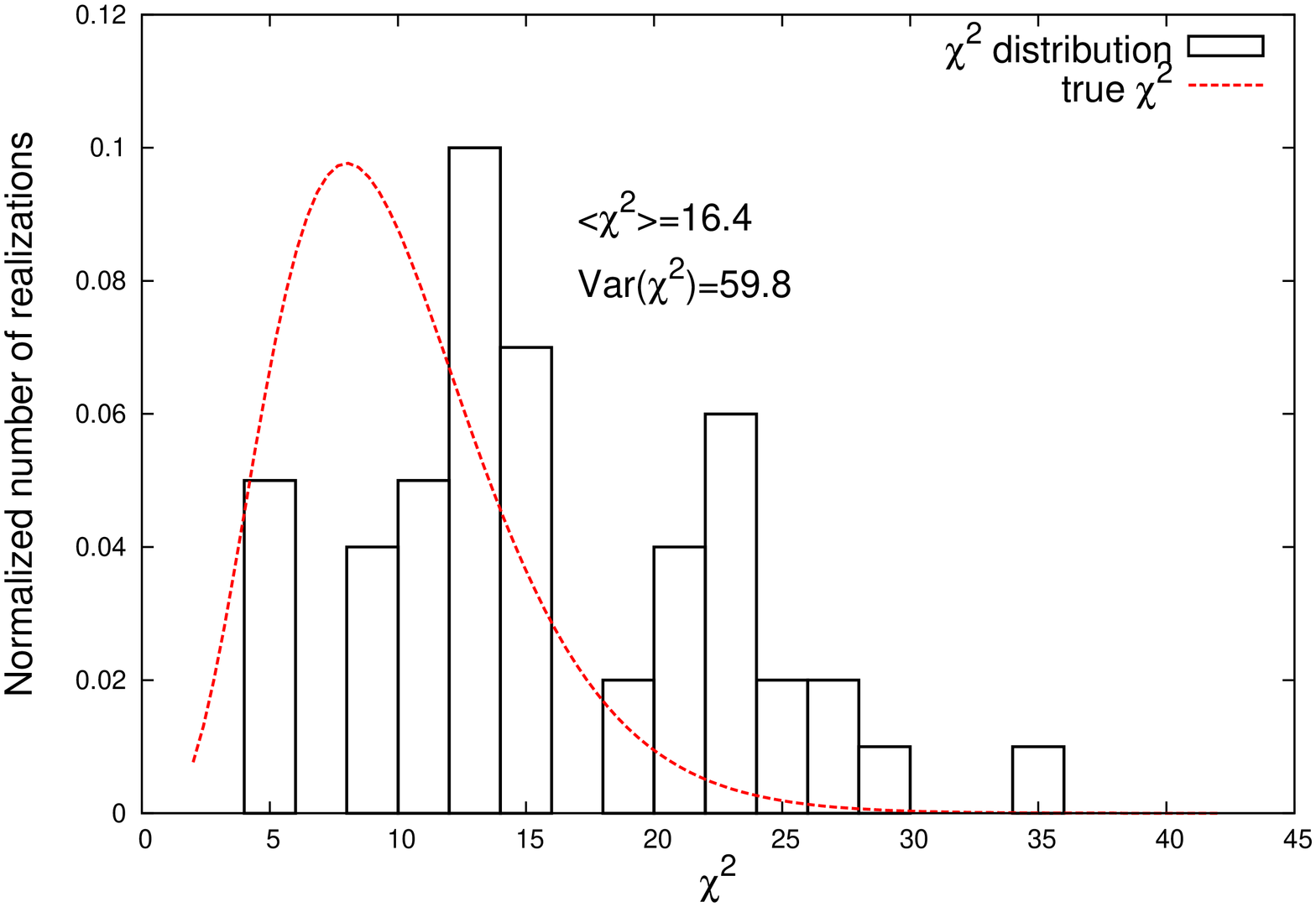}
\caption{Normalized $\chi^2$ distribution based on the
    covariance matrix from the 150 realizations of the fiducial
    cosmology. We consider 10 bins in peak height with equal
    numbers of peaks in each bin (on average). Black histograms represent our data and
    red curve represents a theoretical $\chi^2$ distribution with 10
    degrees of freedom.  The top panel shows the distribution for 150 fiducial
    realizations.  The bottom panel is the distribution for 50 realizations with
    $\OmegaM$ increased by 10\% from its fiducial value.} 
\label{fig:chi2}
\end{figure}

We first examine the distribution of $\chi^2$ values (Eq.~\ref{eq:chi}) in the fiducial cosmology, using the covariance matrix calculated over the 150 realizations and dividing the peak heights into 10 bins of equal numbers of peaks, on average. The top panel of Fig.~\ref{fig:chi2} shows that the distribution observed in the simulations is reasonably well represented by a true $\chi^2$ distribution with 10 degrees of freedom, although with a slight deviation manifest by the somewhat larger variance.

When comparing a modified cosmology to the covariance matrix of the fiducial cosmology, we see that the $\chi^2$ distribution strongly diverges from a true $\chi^2$ law (lower panel of Fig.~\ref{fig:chi2}). This result illustrates the potential of this method to constrain cosmological parameters.  The next step is to compute the constraints we would achieve with the \Euclid\ survey using the Fisher formalism. 

%
%

\subsubsection{Fisher information}

\begin{figure}
\centering
\includegraphics[width=0.3\textwidth,clip,angle=270]{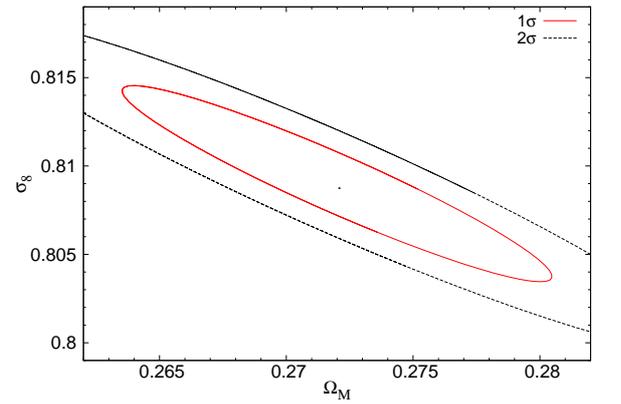}
\caption{Fisher joint conditional constraints on $\OmegaM$ and $\sigma_8$ with non-tomographic peak count statistics for a \Euclid-like survey;
    the red and black contours delineate 1- and 2-$\sigma$ significance limits.}
\label{fig:fishall}
\end{figure}

Following Eq.~(\ref{eq:fish}) we compute the Fisher matrix and invert it to obtain constraints on the cosmological parameters.  Two-dimensional constraints are plotted in Fig.~\ref{fig:fishall} with 1$\sigma$ and 2$\sigma$ confidence contours.  These constraints are summarized in Table~\ref{ultimatetable}.



\section{Tomography}
\label{sec:tomography}
We develop a tomographic approach to peak count statistics with the aim of exploiting the radial information by dividing the source galaxies into redshift bins and detecting peaks to each source plane separately.  We then perform a joint statistical analysis of the multi-plane peak counts.  For example, with two redshift bins we would have a peak distribution consisting of 20 bins in which the first 10 bins represent the distribution of peaks detected to the first source plane, and the remaining 10 the distribution to the second source plane.  Our analysis employs the full covariance of these 20 bins. We note that this approach differs from that employed by \citet{Dietrich+10} in that we do not attempt to localize individual peaks in redshift space; the two approaches, however, access the same information. An important issue with tomography is to ensure that we have enough galaxies in each aperture for the average ellipticity noise to be negligible compared to the average tangential shear.

\subsection{How to slice the redshift dimension}
Adopting a Gaussian random distribution of intrinsic galaxy ellipticities with zero mean and dispersion $\sigma_{\epsilon}$, the shape noise over an aperture is 
\begin{equation}
\sigma_{\rm ap} \approx \frac{\sigma_{\epsilon}}{\sqrt{2N_{\rm ap}}},
\end{equation}
where $N_{\rm ap}$ is the average number of galaxies in the aperture.  We denote $y$ as the desired ratio between the average tangential shear, $\langle \gamma_{\rm t} \rangle _{\rm ap}$, and the aperture shape noise.  An estimate of the number of source galaxies required per aperture is then
\begin{equation}
  N_{\rm ap} \approx \left(\frac{y\sigma_{\epsilon}}{\sqrt{2} \langle \gamma_{\rm t} \rangle _{\rm ap}}\right)^2.
\end{equation}

We take a shape-noise dispersion of $\sigma_\epsilon=0.3$ and an average shear value of 0.04 \citep[e.g., ][]{JT03,JS97}. 
These values with $y=7$ yield a required number of about 1400 source galaxies per aperture.

Working with fixed aperture size, the available number of galaxies per aperture depends on redshift. Using the distribution of redshifts in our simulations (Eq.~\ref{eq:distriz}) normalized to the mean $\Euclid$ galaxy density of 30 galaxies per square arc-minute, we can estimate the number of galaxies per aperture for any slice of redshift. For the most distant redshift bin, the condition of having at least 1400 galaxies per aperture is satisfied for $1.43<z\leq2$.  We adopt the same number of galaxies per redshift slice to avoid favoring any particular redshift bin.  The condition on a minimal number of galaxies per aperture then directly  translates into a condition on the maximum number of redshift bins.  We also do not use the $z\leq0.5$ redshift range to avoid diluting the signal. These conditions allow us to perform a tomographic analysis with up to five redshift slices between redshift 0.5 and 2. We note that relaxing the constraint on the shear to ellipticity ratio would allow more redshift slices. We also note that this approach is limited by the uncertainty on the photometric redshift information, which is on the order of $\sigma(z)=0.05\times(1+z)$.





For a first tomographic study, we use the following five redshift slices with equal numbers of galaxies: $0.5<z\leq0.73$, $0.73<z\leq0.93$, $0.93<z\leq1.15$, $1.15<z\leq1.43$, and $1.43<z\leq2$.  The mean density in each slice is about five galaxies per square arc-minute. This corresponds to about 1400 galaxies per aperture and a shear to ellipticity ratio of about seven.

\subsection{Results}
\begin{figure}
 \centering
 \includegraphics[width=0.35\textwidth,clip,angle=270]{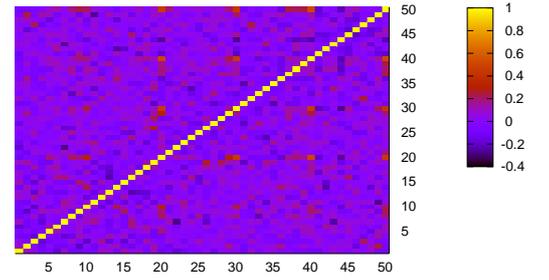} 
 \caption{Correlation matrix of the shear peak distribution for a 100 square-degree field with tomography, calculated by averaging over 150 realizations of the fiducial cosmology (see Eq.~\ref{eq:cov}). Each redshift slice is divided into ten bins of peak height.  This is the covariance matrix normalized to unit diagonal. Correlations between bins are less than 20\%, except at the highest signal-to-noise.}
  \label{fig:correl2}
\end{figure}

We use the same simulations and shape noise realizations as in Section~\ref{sec:stats} when studying the two-dimensional peak counts.  The size of the peak-amplitude bins is determined to include the same number of peaks in each bin of a given redshift slice, based on the mean peak counts in our fiducial cosmology. Fig.~\ref{fig:correl2} shows the full correlation matrix across all bins and source redshift planes, with the first 10 bins corresponding to the lowest redshift source plane and followed in sequence out to the highest redshift plane.  

\begin{figure}
 \centering
 \includegraphics[width=0.35\textwidth,clip,angle=270]{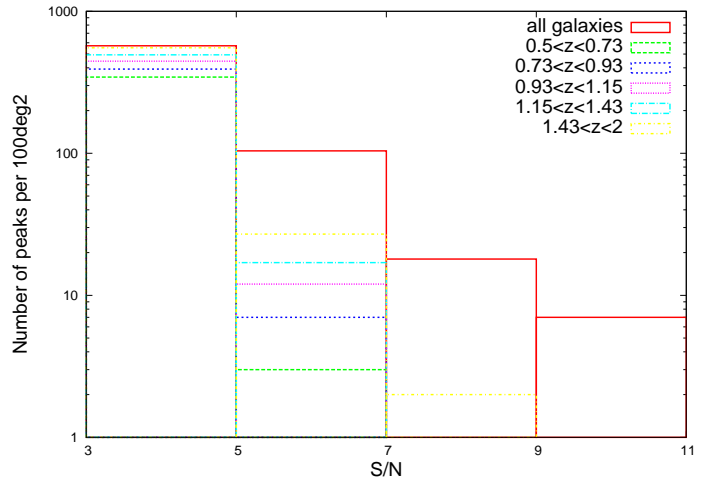} 
 \caption{Mean number of peaks over the 150 realizations of the
   fiducial cosmology. The red histogram corresponds to peaks detected in
   the 2D analysis while green, blue, pink, cyan, and yellow,
   respectively, correspond to peaks detected in the $0.5<z\leq0.73$,
   $0.73<z\leq0.93$, $0.93<z\leq1.15$, $1.15<z\leq1.43$, and
   $1.43<z\leq2$ redshift slices.}
  \label{fig:numberpeaks}
\end{figure}

Table~\ref{tab:numberpeaks} gives the mean and one sigma variation of the number of peaks over the 150 realizations of the fiducial cosmology. The mean number of peaks are also shown in Fig.~\ref{fig:numberpeaks}. 


As seen from Table~\ref{tab:numberpeaks} and Fig.~\ref{fig:numberpeaks}, peaks detected toward a low redshift source plane tend to also be detected when using higher redshift planes.  This correlates the peak counts between source planes, as indicated by the non-zero off-diagonal elements of the correlation matrix, especially in the higher signal-to-noise bins. The bins are not fully correlated, however, because new peaks are detected beyond the lower redshift source planes as we move outward. This tomographic view of the peak distribution contains valuable cosmological information that increases the constraining power of the peak counts.
  
\begin{table*}[ht!]
  \caption{Mean and one sigma variation of the number of peaks over the 150 realizations of the fiducial cosmology.}
\centering
\begin{tabular}{cccccc}
  \hline
  \hline             &  $3<\frac{\rm S}{\rm N}<5$ & $5<\frac{\rm S}{\rm N}<7$ & $7<\frac{\rm S}{\rm N}<9$  & $9<\frac{\rm S}{\rm N}$  & $3<\frac{\rm S}{\rm N}$ \\ 
 \hline 
         All galaxies ($0.5<z\leq2$)  &  572 $\pm$ 19  & 104 $\pm$ 12  & 18 $\pm$ 5 & 7 $\pm$ 3  & 702  $\pm$ 20  \\
         $0.5<z\leq0.73$   & 344 $\pm$  16  & 3 $\pm$ 2  & 0$\pm$ 0  & 0 $\pm$ 0  & 347  $\pm$ 16  \\
         $0.73<z\leq0.93$   & 392 $\pm$ 16   & 7 $\pm$ 3  & 0$\pm$ 1  & 0 $\pm$ 0  & 399 $\pm$ 16   \\
         $0.93<z\leq1.15$   & 445 $\pm$ 18   & 12$\pm$ 4   & 1$\pm$ 1  & 0 $\pm$ 0  & 458  $\pm$19   \\
         $1.15<z\leq1.43$   & 494 $\pm$ 19   & 17 $\pm$ 5  &1 $\pm$ 1  & 0 $\pm$ 0  & 512 $\pm$20    \\
         $1.43<z\leq2$   & 553$\pm$ 16    & 27 $\pm$ 6  &2 $\pm$ 1  & 0 $\pm$ 0  & 583  $\pm$ 17  \\

  \hline 
  \hline

\end{tabular}
\label{tab:numberpeaks}
\end{table*} 
  
\begin{figure}[h]
\centering
\includegraphics[width=0.3\textwidth,clip,angle=270]{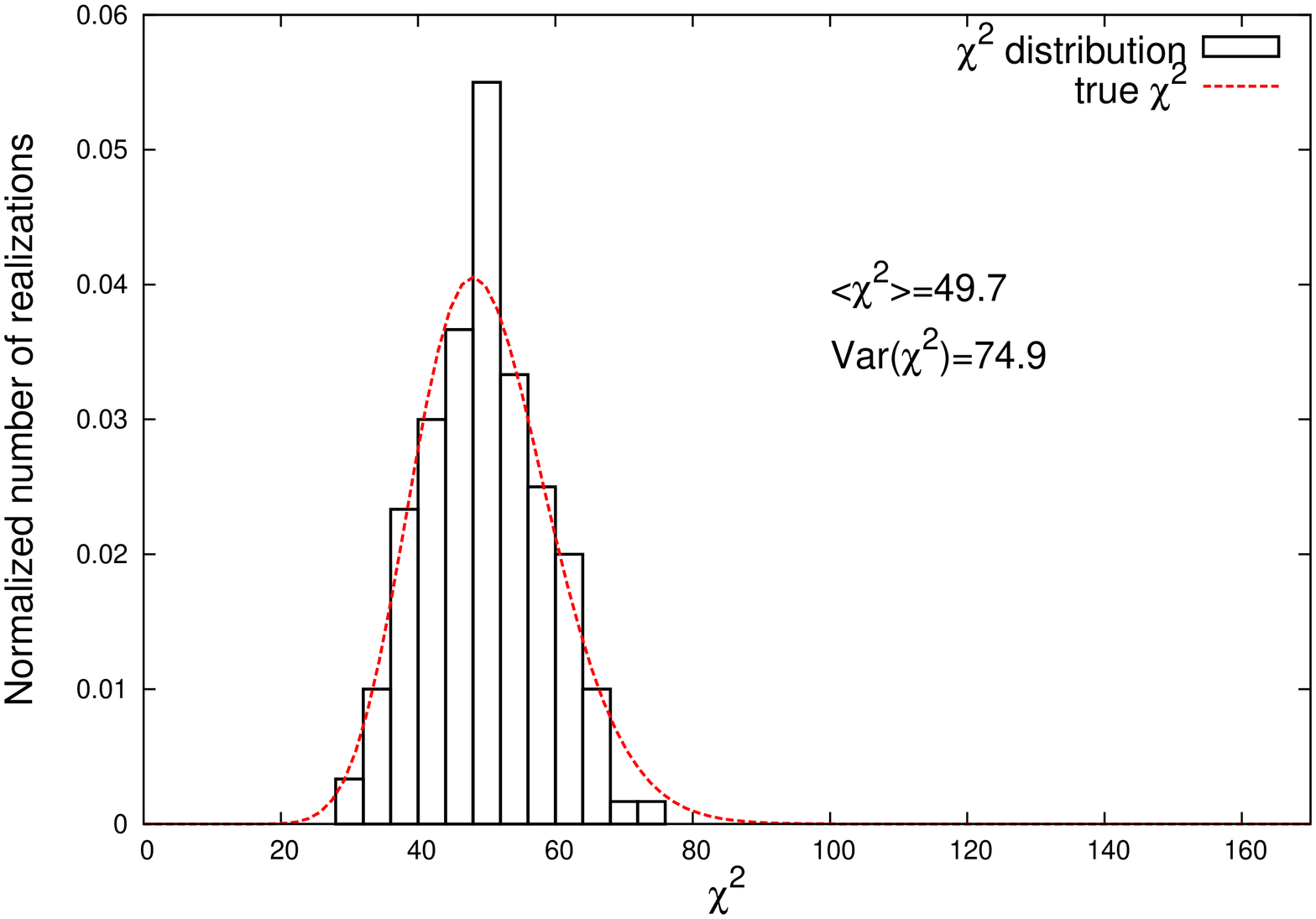} \\
\includegraphics[width=0.3\textwidth,clip,angle=270]{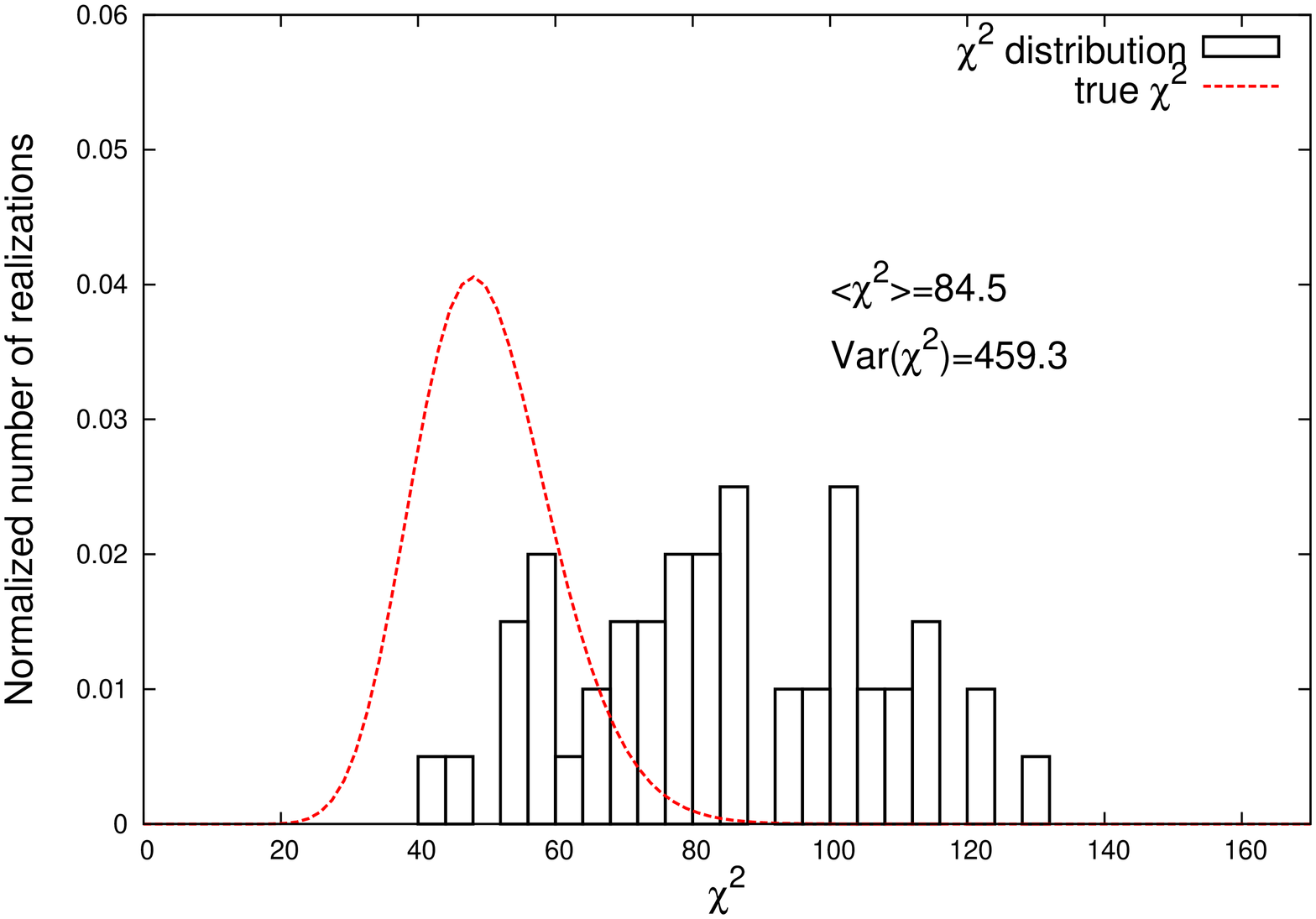}
\caption{Normalized $\chi^2$ distributions with respect to the covariance matrix for 150 realizations of the fiducial cosmology with five source redshift planes.  Each of the five source planes is associated with 10 signal-to-noise bins containing equal numbers of peaks.  Black
histograms trace the distribution observed in the simulations and the red curve represents a theoretical $\chi^2$ distribution with 50 degrees-of-freedom.  The top panel gives the distribution for the 150 realizations of the fiducial model.  The lower panel gives the distribution for the 50 realizations of the alternate cosmology with $\OmegaM$ increased by 10\% from its fiducial value.}
\label{fig:chi2_tomo}
\end{figure}

This can be appreciated from the differences in the $\chi^2$ distributions shown in Fig.~\ref{fig:chi2_tomo}.  The black histogram in the upper panel gives the distribution in the simulations, compared to a true $\chi^2$ distribution with 50 degrees-of-freedom traced by the solid red line.  
The lower panel gives the distribution of our $\chi^2$ variable for the same non-fiducial cosmology considered in the lower panel of Fig.~\ref{fig:chi2}.  As before, the observed histogram strongly differs from the pure $\chi^2$ distribution.  The fact that the histograms differ even more than in the two-dimensional case illustrates our increased ability to distinguish these cosmological models.  


\begin{figure}
\centering
\includegraphics[width=0.3\textwidth,clip,angle=270]{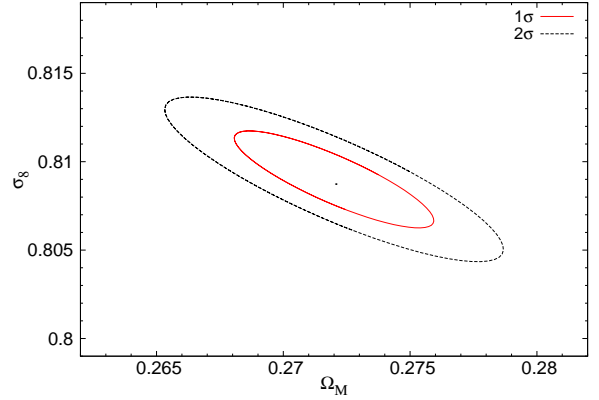}
\caption{Predicted joint conditional constraints on $\OmegaM$ and $\sigma_8$ for a $\Euclid$-like survey using tomographic peak count statistics.  Solid red and dashed black contours correspond to 1$\sigma$ and 2$\sigma$ confidence regions, respectively.}
\label{fig:fishzzz}
\end{figure}

We quantify this gain with a Fisher analysis of the same cosmological parameters considered in the two-dimensional case. The increase in the number of effective degrees-of-freedom in the peak distribution drops the precision on our estimated covariance matrix to 15\%, calculated according to \citet{Taylor+14}. Doubling the number of realizations to $R=300$ would provide 10\% precision, a small gain compared to the computational time required to generate twice as many realizations. We also verify that $R=50$ realizations of the modified cosmologies is sufficient for calculation of the derivatives of the mean counts: as before, they are stable beyond 30 realizations. The constraints from our tomographic analysis are given in Fig.~\ref{fig:fishzzz} and listed in Table~\ref{ultimatetable}, and they are compared to the two-dimensional case in Fig.~\ref{fig:fishzzzall}.

\section{Discussion}
\label{sec:discussion}

\begin{table}[ht!]
\caption{Predicted cosmological parameter constraints for a \Euclid-like survey.  The numbers give $1\sigma$ uncertainties and the corresponding relative percentages of the fiducial parameter values $\OmegaM=0.272$ and $\sigma_8=0.809$.}
\centering
\begin{tabular}{ccc}
  \hline
  \hline             &  $\delta_{\OmegaM}$  & $\delta_{\sigma_8}$  \\ 
 \hline 
      Unmarginalized &&\\
      All galaxies     & 0.0012 (0.43\%)      & 0.0018 (0.22\%)  \\
      Tomography       & 0.0010 (0.35\%)      & 0.0014 (0.17\%)   \\
 \hline
      Marginalized &&\\
      All galaxies     &0.0037 (1.34\%)     &0.0056 (0.69\%)   \\
      Tomography       &0.0018 (0.66\%)     &0.0026 (0.32\%)   \\
  \hline 
  \hline

\end{tabular}
\label{ultimatetable}
\end{table}

\begin{figure}
\centering
\includegraphics[width=0.3\textwidth,clip,angle=270]{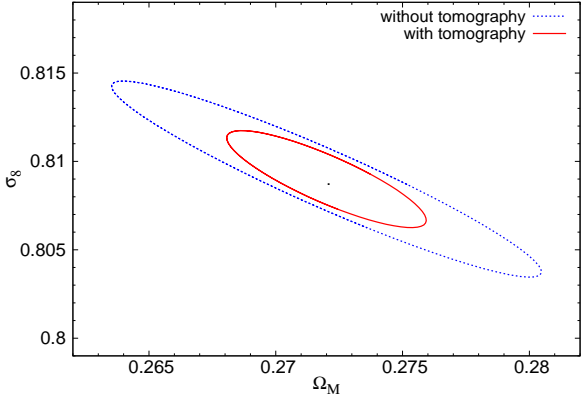}
\caption{Fisher ellipses (at $1\sigma$ ) for $\OmegaM$ and $\sigma_8$ for a $\Euclid$-like survey.  The blue dashed curve shows the joint conditional constraints without tomography, while the solid red contour gives those when using tomography with five redshift bins, demonstrating the important gain.}
\label{fig:fishzzzall}
\end{figure}

A number of authors have investigated the potential of lensing peak counts as a cosmological probe, although few have considered large Stage IV projects such as \Euclid\ \citep{Yang2011,Maturi+11,Hilbert+12,Marian2012,Marian2013}. \citet{Kratochvil2010}, for example, examined the difference in the $\chi^2$ distributions from different cosmological models; however, it is difficult to make a direct comparison with our results since they varied a different set of cosmological parameters. In their pioneering study of tomographic peak counts, \citet{Dietrich+10} considered a CFHTLS-like survey of about 180deg$^2$.

In this work we focus on a typical Stage IV survey characterized by the redshift distribution of a \Euclid-like survey, using a suite of independent numerical simulations to examine possible cosmological constraints. Figure~\ref{fig:fishzzzall} and Table~\ref{ultimatetable} quantify gains in constraining power by using tomography. We improve the marginal constraints on $\OmegaM$ and $\sigma_8$ by more than a factor of two over the two-dimensional (non-tomographic) analysis. As to be expected, the conditional constraints are improved by smaller factors: about 1.2 for $\OmegaM$ and 1.3 for $\sigma_8$.


Among previous studies of peak-statistics, only \citet{Dietrich+10} and \citet{Yang2011} have applied tomography, the former maximizing the peak signal-to-noise given the redshift distribution of galaxies, and the latter placing source galaxies at either $z_s=1$ or $z_s=2$ and ray-tracing through simulations. \citet{Dietrich+10} demonstrated the gain from tomographic peak counts for a survey of 180deg$^2$.  In their LSST-like survey of 20,000deg$^2$, \citet{Yang2011} noted an improvement of the marginal constraints by factors of two to three when using tomography, in qualitative agreement with our results.

We also note that the impact of shape noise is effectively reduced with tomography, in particular for the peaks generated by structures at the higher redshifts.  Binning the sources into redshift planes removes the shape noise contributed by foreground galaxies that do not carry any signal on the higher redshift peaks.

We compare our conditional constraints to those obtained by other authors.  These vary over the range $0.0006<\delta_{\OmegaM}<0.0009$, according to \citet{Hilbert+12} and \citet{Marian2012,Marian2013}, and $0.0013<\delta_{\sigma_8}<0.0016$, according to the same authors and \citet{Maturi+11}. While these studies differ in a number of respects, the agreement on the conditional constraints among these authors and our results is very good.  Indeed, we reach $(\delta_{\OmegaM},\delta_{\sigma_8}) =(0.0012,0.0018)$ without tomography, and $(0.0010,0.0014)$ with tomography. The very small difference from the literature can be attributed to our use of a slightly lower survey area: 15,000deg$^2$ compared to 18,000 to 20,000deg$^2$ in these other studies.

\begin{figure}
\centering
\includegraphics[width=0.5\textwidth,clip]{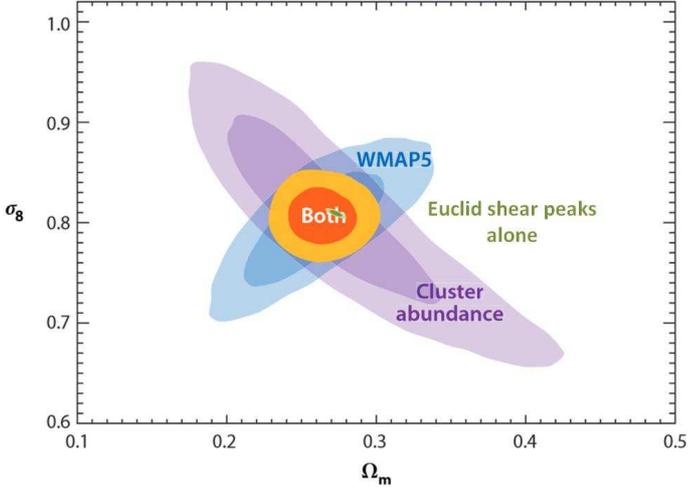}
\caption{Constraints from tomographic peak counts compared to current galaxy cluster constraints, with contours giving the 1- and 2-$\sigma$ confidence limits.  The violet shading represents constraints from the maxBCG cluster catalog, blue those from WMAP-5, and the yellow their combination; they have been adapted from \citet{Allen+11}. The green contours give the tomographic peak-count constraints for a \Euclid-like survey covering 15,000 square degrees.}
\label{fig:compallen}
\end{figure}

Finally, we note that some authors have examined constraints on other cosmological parameters.  In particular, it has been found that shear peaks have a good ability to constrain the dark energy equation-of-state $w$ \citep[e.g.,][]{Yang2011,Hilbert+12,Marian2012,Marian2013} and primordial non-Gaussianity $f_{\rm NL}$ \citep[e.g.,][]{Maturi+11,Hilbert+12}.
We also compare our forecasted constraints with those from other cosmological probes, in particular from cluster abundance studies.  While some lensing peaks do arise from individual clusters, peak statistics represent a more general description of the matter distribution because many originate from projections along the line of sight.  Figure~\ref{fig:compallen} compares our predicted constraints from tomographic peak statistics with those from current galaxy cluster constraints as summarized by \citet{Allen+11}.
The peak counts yield much tighter constraints than those from the current cluster analyses, which is not surprising because we are comparing present day cluster constraints to future lensing counts.

\begin{figure}
\centering
\includegraphics[width=0.5\textwidth,clip]{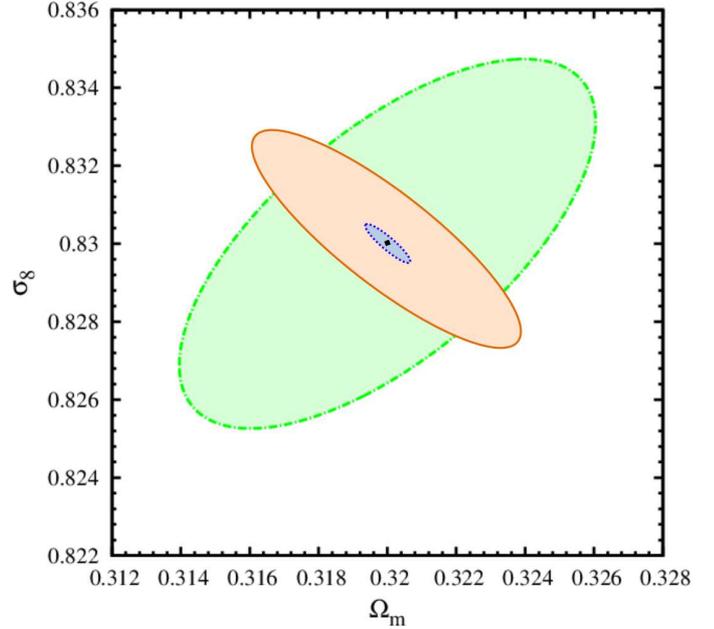}
\caption{Comparison of predicted constraints from \Euclid\ clusters and peak counts for a \Euclid-like survey (1$\sigma$ confidence limits). The orange ellipse traces the tomographic peak-count results. The blue dotted ellipse reports the constraints obtained from cluster number counts with a 3$\sigma$ selection function assuming a perfectly known observable-mass relation \citep[see][]{Sartoris+15}. The green dash-dotted ellipse shows the same cluster constraints but leaving the 4 scaling relation parameters (bias, scatter, and their evolution) completely free to vary. We note the change of scale and shift of fiducial parameter values from Fig.~\ref{fig:compallen}.}
\label{fig:compeuclid}
\end{figure}

A more pertinent comparison is between constraints predicted from cluster photometric sample for the $\Euclid$ survey and the peak count constraints.  This is shown in Fig.~\ref{fig:compeuclid}. The cluster constraints have been evaluated by \cite{Sartoris+15} for a \Euclid\ cluster catalog considering the information provided by cluster number counts. We note that, unlike what has been done in \citet{Sartoris+15}, constraints from clusters have been performed by varying only the $\sigma_8$ and $\OmegaM$ cosmological parameters and the 4 parameters that describe the bias, the scatter of the observable mass relation, and their redshift evolution. This has been done to compare in a more appropriate way the constraints obtained from the clusters and those from the shear peaks. The reduction in the number of free cosmological parameters explains why the cluster constraints shown in Fig.~\ref{fig:compeuclid} are smaller than those shown in \cite{Sartoris+15}. We see that the constraints from tomographic peak counts are weaker than cluster counts by an order of magnitude when supposing that the observable-mass relation is fully known a priori (blue dotted ellipse). However, the shear-peak constraints are almost twice as strong when not making any assumption on the scaling relation parameters and their evolution (green dash dotted ellipse). In addition, it is worth noting that the shear-peak contours are orthogonal to the clusters when the observable-mass relation is not known a priori. This essentially shows the value of using both clusters and shear-peak statistics. As it is difficult to predict how well we will be able to constrain the scaling relation, we show here two extreme cases for the cluster constraints, with the idea that the observational constraints should lie somewhere between the blue and green ellipses. In particular, the green contours are very conservative, as we could in principle already constrain the scaling relation at redshift $z=0$, which would reduce the errors on the cosmological parameters.

We expect peak statistics to complement more standard two-point lensing statistics.  In fact, the study by \citet{Dietrich+10} suggests that peak statistics could yield tighter constraints than the classical 2-point correlation function, a result expected given that the peak statistics contain higher order correlations, and later confirmed by, e.g., \citet{Marian2012,Marian2013}. Peak counts would also appear to be less affected by shape measurement systematics than the shear power spectrum in the sense that it is more difficult to reproduce the pattern necessary for peak identification than to affect the amplitude of two-point correlations.  In similar vein, we expect that their respective sensitivities to photometric redshift uncertainties will not be the same.  Overall, the two methods will not share the same systematics and therefore offer important complementarity.  

\section{Conclusion}
We have found that shear peak statistics offer a potentially powerful cosmological probe, in agreement with previous studies. As an advance along these lines, our results clearly illustrate the gain of using tomography in the framework of Stage IV dark energy surveys, i.e., separating the source galaxies into redshift planes and counting peaks to each plane.  With tomography, we improve the conditional (respectively, marginal) constraints by a factor of 1.2 (resp. 2) on $\OmegaM$ and $\sigma_8$.

For a large-area survey, typified here as that from a \Euclid-like mission, we estimate that the peak-count constraints are an order of magnitude less powerful than those predicated from galaxy cluster evolution when the observable-mass relation is fully known a priori, while they are twice as strong when not making any assumption on this relation. The peak counts, however, have the great advantage of not relying on such a scaling relation that may prove difficult to establish to high accuracy.
 
We have only explored the two parameters $\OmegaM$ and $\sigma_8$ in the present study, but plan to extend to other parameters, including the dark energy equation-of-state and primordial non-Gaussianity in future work. Further topics warranting exploration include the impact of various systematics, such as intrinsic alignments, photometric redshift errors, and shape measurement errors.  These additional studies will quantify the extent to which peak counts are complementary to cosmic shear measurements.

The primary technical challenge in application of peak counts is the production of large suites of numerical simulations to calculate both the expected mean number of peaks and their covariance matrix over the cosmological parameter space.  It is not, however, unique to the counts: all lensing studies face the same challenge because valuable signal, even in the two-point statistics of cosmic shear, originates in the non-linear regime.  We therefore expect peak counts to accompany the more standard lensing measures in application to large lensing surveys.  

\begin{acknowledgements}
We are grateful to the anonymous referee for his/her careful reading and helpful and constructive comments that improved the paper. NM thanks the {\it Ecole Normale Superieure de Cachan} and the {\it Laboratoire AstroParticle et Cosmologie} for financial support in the early stage of this work. BS acknowledges financial support from MIUR PRIN2010-2011 (J91J12000450001), from the PRIN-MIUR 201278X4FL grant, from a PRIN-INAF/2012 Grant, from the “InDark” INFN Grant and from the “Consorzio per la Fisica di Trieste”. A portion of the research described in this paper was carried out at the Jet Propulsion Laboratory, California Institute of Technology, under a contract with the National Aeronautics and Space Administration.
\end{acknowledgements}

\bibliographystyle{aa}
\bibliography{shear_peaks}
\end{document}